\documentclass[pre,twocolumn,showpacs,preprintnumbers,amsmath,amssymb]{revtex4}

\usepackage{graphicx}
\usepackage{amssymb}

\begin{document}



\title{Dynamics and structure of an aging binary colloidal glass}


\author{Jennifer M. Lynch}
\author{Gianguido C. Cianci}
\author{Eric R. Weeks}
\email{weeks@physics.emory.edu} 

\affiliation{Department of Physics, Emory
  University, Atlanta, GA 30322, U.S.A.}

\begin{abstract}
  We study aging in a colloidal suspension consisting of
  micron-sized particles in a liquid.  This system is made
  glassy by increasing the particle concentration.
  We observe samples composed of
  particles of two sizes, with a size ratio of $1:2.1$ and a
  volume fraction ratio $1:6$, using
  fast laser scanning confocal microscopy.  This technique
  yields real-time, three-dimensional movies deep inside the
  colloidal glass.  Specifically, we look at how the size,
  motion and structural organization of the particles relate to
  the overall aging of the glass.  Particles move in spatially
  heterogeneous cooperative groups.  These mobile regions tend
  to be richer in small particles, and these small particles
  facilitate the motion of nearby particles of both sizes.
\end{abstract}


\maketitle

\section{Introduction}

A liquid can be crystallized by slowly lowering
its temperature below $T_{m}$, the crystal melting
temperature. On the other hand a rapid temperature quench
to below $T_{g}$, the glass transition temperature, yields
an amorphous solid with interesting non-equilibrium properties
\cite{sillescu99,gotze92,EdigerAngellNagelJPhysChem1996,DawsonCurrOpColloidInterfSci2002,StillingerScience1995,AngellScience1995,angell00}.
More specifically, the glassy dynamics slow down dramatically and do
so on macroscopically large time scales.  This phenomenon is known
as aging and has been seen in a variety of glass-forming materials
\cite{hodge95,vanmegenpre98,CianciWeeks2006,CianciWeeksSSC2006,angell00,nowak98,cipelletti02,cipelletti03,cipelletti03b,cipelletti05,leheny04,bonn07,munch03,schweizer07a}.
However, the microscopic mechanisms by which a system ages are
still unclear \cite{ngai07pm}.

Dense colloidal suspensions are good models to study glassy systems.
These are composed of small micron-sized solid particles in
a liquid.  In these systems, the glass transition is reached
by increasing the volume fraction, and a glass is formed once
the volume fraction is greater than $\phi_{g} \approx$ $0.58$
\cite{PuseyvanMegenNature1986}.
Previous work looked at aging in colloidal
glasses composed of particles of one size
\cite{CourtlandWeeksJPhysCondMat2003,CianciWeeks2006}.
A small ($5\%$) size polydispersity was used to
inhibit crystallization on experimental time scales
\cite{KegelLangmuir2000,citeulike:2423268}.  In this manuscript, we
study aging in a binary colloidal system.  To an extent the binary
system facilitates comparison with prior simulations
\cite{KobBarratPRL1997,KobTartagliaJPhysCondMat2000}, but our
larger size ratio ($2.1 : 1$) also highlights the role of the
different sizes in the aging process.

In particular, we find that smaller particles are more mobile
than expected. These significantly larger movements are
key components of the aging process since, in our system,
the structural arrangement of the colloidal particles
completely determines the ``age'' \cite{CianciWeeks2006}.
Particles moving to new positions change the structure and
allow the system to become ``older''.  Thus, we conjecture
that the small, unusually mobile particles are a key component
of the aging process, and perhaps facilitate other, more
subtle structural changes of the slower moving large particles
\cite{hoffman92,mewis94,vanmegen01}.  This is reflected in the
tendency for the neighbors of small particles to have higher
mobility on average, and gives insight into the structural
origin of the aging process.

\section{Experimental Methods}
\label{methods}

\begin{figure}[b]
\centering
\includegraphics[width=7cm]{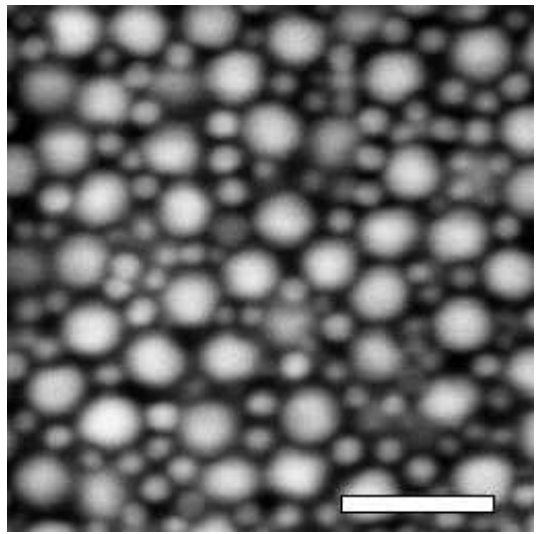}
\caption{Confocal micrograph of the binary colloidal glass studied
in this experiment. The image was taken $20 \mu m$ inside the
sample.  The scale bar represents $5 \mu m$.
}
\label{pic}
\end{figure}

We use two sizes of poly(methyl methacrylate)
(PMMA) particles that are dyed with Rhodamine 6G
\cite{DinsmoreWeitzApplOptics2001,AntlCollSurf1986}, as shown
in Fig.~\ref{pic}.  These particles are suspended in a mixture
of $85\%$ cyclohexylbromide and $15\%$ decalin by weight.
This mixture closely matches both the density and refractive
index of the particles \cite{DinsmoreWeitzApplOptics2001}.
The small particles have a radius of $a_{S} = 0.56 \pm .05 \mu m$
and the large particles have a radius of $a_{L} = 1.17 \pm .05 \mu
m$; the error bars reflect the uncertainty of the mean radius
of each particle species, and additionally each species has a
polydispersity of $\sim 5$\%.  The number ratio of small to large
particles is $1.48 : 1$, but given the large size ratio $1 : 2.1$,
the volume fraction ratio is $1 : 6$.  The total volume fraction
is $\phi_{S+L}\approx 0.62$.  We do not see any demixing of the
two colloidal species, perhaps because the sample is glassy and
thus particle motion is difficult.  

In individual {\it monodisperse} samples, the two particle
species would each be expected to have a glass transition at
$\phi_g \approx 0.58$.  For volume fractions slightly less
than $\phi_g$, a monodisperse sample should form crystals in
a finite time, and for samples with volume fractions slightly
above $\phi_g$, previous colloidal samples similar to ours
only find crystals which nucleate at flat boundaries and
which only grow into the bulk of the sample a short distance
\cite{WeeksWeitzScience2000,PuseyvanMegenNature1986,henderson96}.
It is the lack of homogeneous crystal nucleation that is used to
define $\phi_g$ for these monodisperse samples, although in practice
it is probable that a slight polydispersity of the particles is
needed to truly prevent crystallization \cite{henderson96,auer01b}.
Given that the two monodisperse colloidal samples should have the
same $\phi_g$, it is worth noting that our binary sample is not
analogous to mixtures of two polymer glasses with different glass
transition temperatures $T_g$, which are known to have interesting
behaviors \cite{Cowie1989,Cowie2005}.  In practice, for our sample,
we cannot use the lack of crystallization to define $\phi_g$ as
our binary sample does not crystallize at any volume fraction.
Instead, we regard our sample as glassy as the dynamics do not
equilibrate, but change with age for as long as we have observed,
as will be discussed in detail in Sec.~\ref{results}.

We use fast laser scanning confocal microscopy which yields clear
images deep inside our dense samples \cite{prasad07a}.  Despite the
high density, the two colloidal species can be easily discerned.  We
acquire three-dimensional scans of our sample yielding a $53 \times
53 \times 15$~$\mu$m$^3$ observation volume.  The acquisition time
for one 3D image ($\sim 8$~s) is several orders of magnitude faster
than the diffusion time for the particles at this volume fraction.
We continuously follow the sample's aging over $1.5$ hours at a rate
of three images per minute.  This allows us to track $\sim 5000$
particles in three dimensions over the course of the experiment
\cite{CrockerGrierJCollInterfSci1996,DinsmoreWeitzApplOptics2001}.

To study aging, we first must initialize aging in the sample.
There are two approaches to initialization in any glassy
system \cite{mckenna03}.  The first is to quench a sample from a
liquid state.  In a conventional glass, this is done by rapidly
lowering the temperature or increasing the pressure.  In a hard
sphere or colloidal glass, this would be done by starting with a
sample with a volume fraction $\phi < \phi_{g}$ and then rapidly
increase the volume fraction to above $\phi_{g}$.  Currently,
there is no experimental technique that allows for this method
in a sample of hard PMMA colloids.  The second method for
initializing aging involves shear rejuvenation:  by applying
a large shear to a system (above any yield stress), in some
cases the aging appears to be reinitialized.  This is the method
used in this study and previous studies on aging with colloids
\cite{CourtlandWeeksJPhysCondMat2003,CianciWeeks2006,cloitre00,bonn02,viasnoff02}.
There are questions as to whether these two methods lead to the
same glass \cite{mckenna03}, but for our current study, we stress
that only the second of these methods is experimentally available.
Experimentally, this leads to a reproducible initial state
\cite{CourtlandWeeksJPhysCondMat2003} and the dynamics appear
similar to a conventional glass aging via a rapid quench from a
liquid state, as will be shown below.

To shear rejuvenate the sample, we begin each experiment by pulling
a small metal wire that is present in the samples with a hand
held magnet, thus shear melting the glass and re-initializing the
aging process.   We begin taking data, and thus define our initial
time $t_{w}=0$, within 20~s after we finish pulling the wire.
In practice, transient flows continue for a few seconds after
cessation of the stirring, and these flows likely still shear the
glass, resulting in some uncertainty of the initial time.  However,
the 20~s delay between ending the stirring and starting the data
acquisition is long enough to allow for these flows to decrease
quite significantly, and we see no signs of shearing even at the
start of data acquisition.  Thus there is at most an uncertainty
of 20~s for our choice of $t_w=0$, and in practice the results
shown in this work are for time scales much longer than 20~s.

\section{Results}
\label{results}

\begin{figure}[b]
\centering
\includegraphics[width=7.0cm]{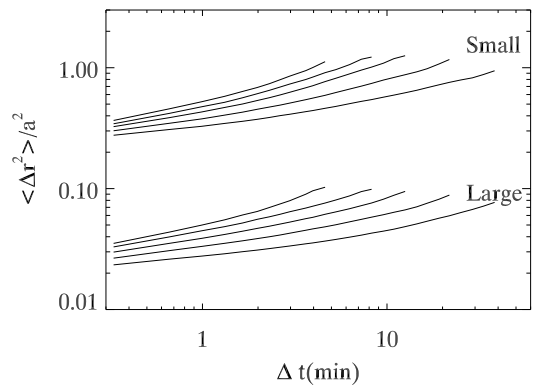}
\caption{Aging mean squared displacement for large and
  small particles of a binary colloidal glass at
  $\phi\approx0.62$. The five curves represent five different
  ages of the sample.  The geometric mean ages of the curves,
  from top to bottom, are $\bar{t}_{w} = 8.7, 15, 26, 45$, and
  $76$ min; see Table~\protect\ref{table1} for details.
  The mean square displacements are normalized relative to each
  particle size, that is, for the small particles we plot $\langle
  \Delta r^2 \rangle / a^2_S$ and for the large particles we plot
  $\langle \Delta r^2 \rangle / a^2_L$.}
\label{msd}
\end{figure}

Aging in glasses is defined as the dependence of the dynamical
properties of the system on the time elapsed since vitrification. To
quantify these changing dynamics, we calculate the mean
squared displacement (MSD), defining displacements $\Delta \vec
r_i(t_w,\Delta t) = \vec{r}(t_w+\Delta t) - \vec{r}(t_w)$, and the
MSD as $\langle \Delta r^2 \rangle_{i,t_w}$.  The angle brackets
indicate an average over all particles $i$ (of a given particle
size), and over all times $t_w$; the MSD is thus a function of the
lag time $\Delta t$.  In practice, we wish to understand the $t_w$
dependence of the MSD as well.  Thus, while we continuously take
data for 1.5~h, we divide the data into five temporal windows,
and calculate the MSD restricting the average over times $t_w$
within each window.  The windows are formed logarithmically based
on the sample age $t_{w}$ and each temporal window is
characterized by its geometric mean age $\bar{t}_w$;
see Table~\ref{table1} for details.
For each window, the MSD is calculated for lag times $\Delta t$
up to $\sim 0.5 \bar{t}_w$, and it is important to recognize that
for the largest lag times in each window, the data are aging
during the observation window.  Thus, care must be taken in
interpreting the MSD curves.  Nonetheless, this method provides a way to
characterize the dynamics for a given age $\bar{t}_w$.

Figure~\ref{msd} shows the MSDs of the glass over a total period
of 1.5 h for the large and small particles.  At short time
scales, the motion of a particle is expected to be diffusive as
each particle explores its local environment without noticing
its neighbors. However, at these high densities, particles are
crowded, and this diffusive regime can only be observed on time
scales shorter than our image acquisition rate and is therefore
not visible in Fig.~\ref{msd}. Instead, we observe a plateau for
most of our time scales, indicating that a particle's motion
is slower and more inhibited than normal diffusion.  This is
because particles are trapped in cages formed by their neighbors
\cite{weeksweitzprl02}.  At long time scales, the MSDs
show an upturn, with the time scale of this upturn increasing with
the age of the sample.

\begin{table}
\caption{\label{table1}
We divide the data into five temporal windows, each window
beginning at $t_w = t_1$ and ending at $t_w = t_2$, with values
$t_1$ and $t_2$ listed below.  For each window, $t_2 = 1.7 t_1$
and the ``age'' $\bar t_w$ is defined as the geometric mean of $t_1$
and $t_2$.  All times are in minutes.
}
\begin{ruledtabular}
\begin{tabular}{ccc}
$\bar t_w$ & $t_1$ & $t_2$ \\
\hline
8.7  &  6.7  & 11.3\\
15  &  11.3  & 20\\
26  &  20    & 34\\
45  &  34    & 58\\
76  &  58    & 100\\
\end{tabular}
\end{ruledtabular}
\end{table}

Both the large and small particles' dynamics are aging similarly,
showing upturns at similar lag times.  Clearly in our experiments,
the aging of each species is strongly coupled to that of the other.
Were we comparing two monodisperse samples with different particle
sizes $a_S$ and $a_L$, the horizontal axis of Fig.~\ref{msd}
would need to be rescaled by the diffusion time scale for each
particle, $\tau_D = 3 \pi \eta a^3 / k_B T$, where $\eta$ is the
solvent viscosity, $k_B$ is Boltzmann's constant, and $T$ is the
absolute temperature.  $\tau_D$ thus depends sensitively on particle
radius $a$, and the ratio of diffusion time scales for our particle
sizes is $(a_L/a_S)^3 \approx 9$.  However, in our data, clearly
the upturns in the MSD curves occur at quite similar time scales
(Fig.~\ref{msd}), rather than differing by an order of magnitude.
These essentially similar upturns indicate the system has a single
time scale for structural rearrangements of both species, and this
time scale grows as the sample ages.  As noted in
Sec.~\ref{methods}, this is perhaps reasonable for a binary
colloidal mixture but stands in contrast to mixtures of polymers
with differing glass transition temperatures
\cite{Cowie1989,Cowie2005}.  In fact, it suggests that in
our experiment the relevant time scale determining structural
rearrangements is $\bar t_w$ (which is the same for both species)
and not the diffusion time scale $\tau_D$ (which differs by a
factor of 12).

In Fig.~\ref{msd} we nondimensionalized $\langle \Delta
r^2 \rangle$ to better compare the behavior of large and
small colloids.  The height of the plateau of the MSD curve
is related to the cage size \cite{weeksweitzprl02}, which would be
the same nondimensional size for two monodisperse samples.
In our binary sample, we observe the nondimensional cage
size is larger for small particles. For example, at $\bar{t}_w =
76$~min and using a time scale $\Delta t=10$~min, we find
$(\langle \Delta r_L^2 \rangle)^{1/2} = 0.25 \mu m = 0.21 a_L$, and
$(\langle \Delta r_S^2 \rangle)^{1/2} = 0.43 \mu m = 0.76 a_S$.  Thus,
the small particles are more mobile in their cages, both in an
absolute sense, and especially so when taking into account their
smaller size.

\begin{figure}[b]
\centering
\includegraphics[width=7.0cm]{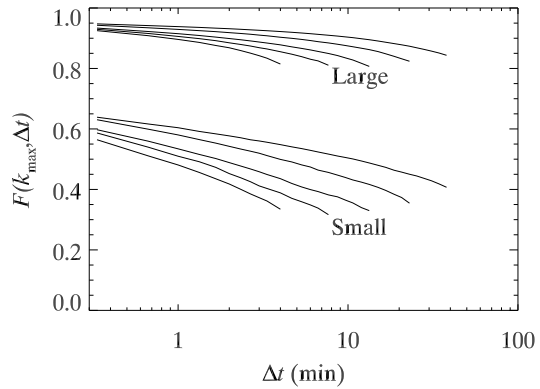}
\caption{Self intermediate scattering function for each
  particle species.  Each group of five curves represent five
  different ages of the sample.  The geometric mean ages of the
  curves, from bottom to top, are $\bar{t}_{w} = 8.7, 15, 26,
  45$, and $76$ min; see Table~\protect\ref{table1} for details.
  Each scattering function is computed for the wave vector $k_{\rm
  max}$ which maximizes the structure factor for that particle
  species.  The specific values are $k_{\rm max}=5.61$~$\mu$m$^{-1}$
  for the small particles and $k_{\rm max}=3.14$~$\mu$m$^{-1}$
  for the large particles.
  }
\label{scattering}
\end{figure}

A further way to characterize the dynamics is to examine the
self intermediate scattering function for the two species.
This is calculated as
\begin{equation}
F_s(\vec k,\Delta t) = \langle \exp(-i \vec k \cdot \Delta \vec
r_i)_{i,t_w}
\rangle
\end{equation}
where the angle brackets again indicate an average over starting
times $t_w$ and over all particles of a given size, similar to
the definition of the mean square displacement.  In practice, we
choose $| \vec k |$ from the maximum of the static structure factor
for each species, and additionally we average over orientations
of $\vec k$.  The results are shown for the five time windows
in Fig.~\ref{scattering}.  $F_s(\Delta t)$ has a downturn for
both species around similar values of $\Delta t$ for each age
$\bar t_w$.  The overall magnitude of $F_s(\Delta t)$ is larger for
the large particles, reflecting that they move less, as shown in
Fig.~\ref{msd}.  Thus the two particle sizes clearly have different
albeit connected dynamics.  As with the MSD curves, note that at
the largest time scales $\Delta t$ shown in Fig.~\ref{scattering},
the sample is aging over that time scale.  Given that the sample
ages before complete decay can be seen, it is difficult to
characterize the decay of $F_s(\Delta t)$, but we note that the
downturn seen at longer lag times is linked to the small
irreversible structural rearrangements of the sample.

\begin{figure}[t]
\centering
\includegraphics[width=9.0cm]{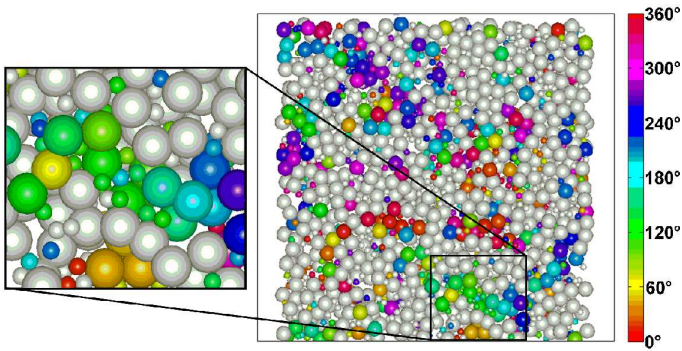}
\caption{
(Color online)
A narrow slice of a three-dimensional sample with width
$\Delta z = 4$~$\mu$m at $t_{w} = 20$~min.
Particles with the $30\%$ greatest displacements $\langle \Delta
r \rangle \geq 0.375$~$\mu$m for $\Delta
t = 10$ min are colored with color indicating the direction
of displacement projected onto the $xy$-plane, where $0^\circ$
indicates motion to the right and $90^\circ$ indicates motion
directly upwards.  When viewed in grayscale, lighter shades indicate
motion upwards, and darker shades indicate motion downwards.
}
\label{matlab}
\end{figure}

We wish to investigate the relationship between structure
and dynamics, that is, how a particle's local environment
influences its motion and the motion of its neighbors
\cite{CianciWeeks2006,CourtlandWeeksJPhysCondMat2003}.
Prior work indicated that the particles rearrange in
cooperative groups \cite{CourtlandWeeksJPhysCondMat2003},
similar to what is seen in supercooled liquids
\cite{EdigerAngellNagelJPhysChem1996,WeeksWeitzScience2000,KegelvanScience2000,DonatiGlotzerPRL1998,harrowell04,sillescu02,berthier04,yamamoto98,ediger00}
and granular materials \cite{Mehta2008,GoldmanSwinney2006}.
A starting point to look for cooperativity
is to ask where the mobile particles are located
\cite{DonatiGlotzerPRL1998,berthier04,CourtlandWeeksJPhysCondMat2003};
Fig.~\ref{matlab} highlights the most mobile particles at
$t_w = 20$~min, using $\Delta t=10$~min.
The coloring indicates the direction of
motion.  Several features are seen.  First, mobile particles
are clustered, as was reported in monodisperse samples
\cite{CourtlandWeeksJPhysCondMat2003}.  A mobile particle is
likely to have mobile neighbors.  Second, particles of similar
colors are likely to be neighbors, indicating that groups of
mobile particles are moving in similar directions.  For example,
the magnified region in Fig.~\ref{matlab} shows a group of particles
all moving approximately up and to the left (green and blue
colors).
Third, occasionally
neighboring particles have significantly different directions of
motion, similar to the ``mixing'' particles seen in supercooled
fluids \cite{weeksweitzprl02}.  This tendency is enhanced
for the small species, which can often move in directions different
from their larger neighbors.  These three observations highlight
that indeed the motion of a particle is related to the motion
of its neighbors, as discussed before.  It is these cooperative
motions that are responsible for the upturn in the MSD curves
(Fig.~\ref{msd}).

\begin{figure}[t]
\centering
\includegraphics[width=7.0cm]{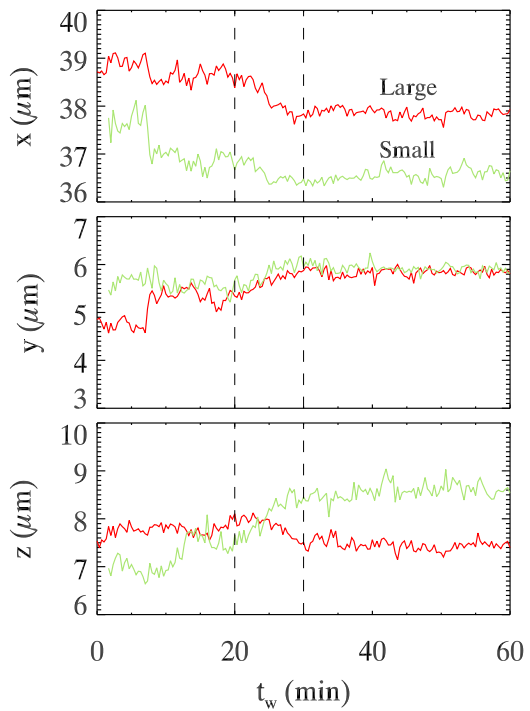}
\caption{
(Color online)
Trajectories of two neighboring
particles:  a large particle (red/dark gray) and a small particle
(green/light gray).  The particles have large displacements at
$t_w=20$~min, using $\Delta t=10$~min.  The vertical dashed
lines indicate this time range $t_w,t_w+\Delta t$.
}
\label{traj}
\end{figure}



The trajectories of two mobile particles are shown in
Fig.~\ref{traj}, further demonstrating the cooperative motion.
These are neighboring particles taken from those shown in
Fig.~\ref{matlab}.  One is a large particle (red/dark gray) and
the other is small (green/light gray).  At $t_w=20$~min they both
begin to move in similar directions, with the motion essentially
complete by $t_w=30$~min.  Subsequent to the rearrangement, these
two particles remain caged for the next 30 minutes (further data
not shown).  Figure \ref{traj} shows that these large displacements
happen slowly in our aging glass; while the particles each move
$\sim 1$~$\mu$m, this takes the whole time interval $\Delta
t=10$~min.  Occasionally some particles move more rapidly, and
some move more slowly; similar results to Fig.~\ref{matlab} can
be seen using other choices of $\Delta t$, and indeed motions of
other durations can be seen in the trajectories of Fig.~\ref{traj}
at other times.

\begin{figure}[t]
\centering
\includegraphics[width=8.0cm]{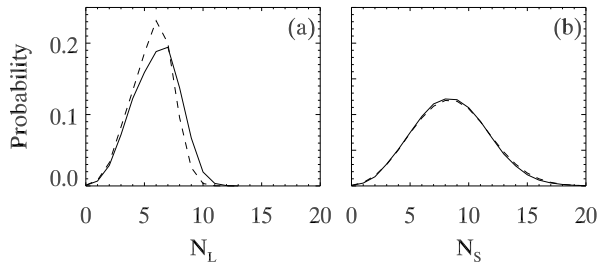}
\caption{Probability of a large particle (solid line) and a small
  particle (dashed line) having a certain number of (a) large
  neighbors and a certain number of (b) small neighbors.  These
  distributions do not change as the sample ages.}
\label{prob}
\end{figure}

\begin{figure}[b]
\centering
\includegraphics[width=7.0cm]{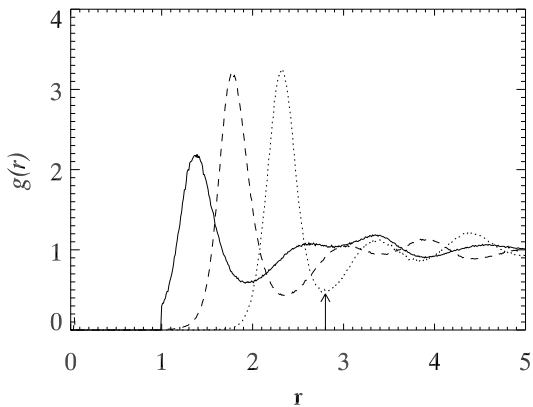}
\caption{Pair correlation functions
  $g_{SS}(r)$ (solid line), $g_{LL}(r)$ (dotted line), and $g_{SL}(r)$
  (dashed line).  The arrow at $r=2.8$~$\mu$m indicates the cutoff used to define nearest neighbors.
  These functions do not change as the sample ages.}
\label{gr}
\end{figure}

These observations do not yet show
how the local structural environment influences the motion.
To quantify the structure of the local environment of a
particle, in Fig.~\ref{prob} we plot the probability of a
particle having $N_{L}$ large neighbors (a) or $N_{S}$ small
neighbors (b). The probabilities are computed separately for
large and small reference particles. We define a neighbor as a
particle within 2.8~$\mu$m from the reference particle being
considered.  This cut-off distance is set by the first minimum
of the pair correlation function for large particles, as shown
in Fig.~\ref{gr}.  However, the trends seen in the distributions
of Fig.~\ref{prob} are not sensitive to small variations
of this value.  Consider Fig.~\ref{prob}(a) which shows the
probability of having $N_{L}$ neighbors for both species of
particles. The two distributions are very similar
indicating that the neighborhoods of both large and
small reference particles contain similar numbers of large
particles. Figure~\ref{prob}(b) shows a similar plot counting
the likelihood of having $N_{S}$ small neighbors for both
species. Again we find comparable distributions, showing that
the sample is statistically homogeneous and there is no size
segregation.  The most probable neighborhood [the maximum of
the probability distribution $P(N_L,N_S)$] is $N_L=7, N_S=10$
for large particles, and $N_L=6, N_S=10$ for small particles.

\begin{figure}[b]
\centering
\includegraphics[width=8.0cm]{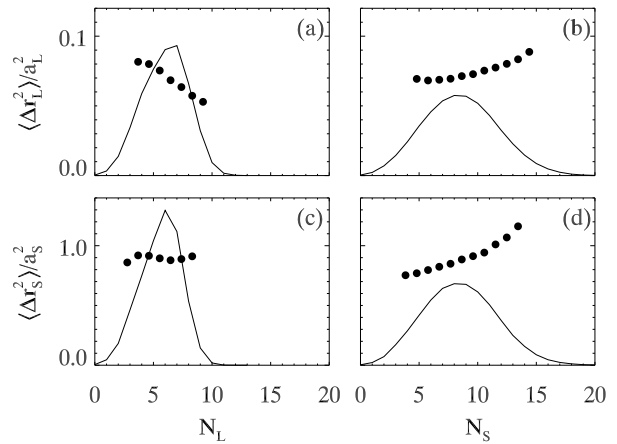}
\caption{Average particle mobility versus the number and type of
  neighbors the particle has (circles) for $\Delta t =$ 10 min.
  The solid line is proportional to the probability of having
  that number and type of neighbors.  (a) Large particles with
  large neighbors, (b) large particles with small neighbors, (c)
  small particles with large neighbors, (d) and small particles
  with small neighbors.  The trends in these graphs do not change
  as the sample ages or with choice of $\Delta t$.  The curves are
  truncated where the probabilities of having that many neighbors
  drops below 3\%.}
\label{mob}
\end{figure}

The probability distributions in Fig.~\ref{prob} do not depend
on sample age.  This is in agreement with previous work on
aging monodisperse glasses where other local measures of
structure did not show any aging \cite{CianciWeeks2006}.  However,
in that study the structure showed some correlation to the
dynamics. To establish whether this still holds in a binary
suspension, we first calculate the mobility of each particle.
To do this, we choose a fixed lag time $\Delta t = 10$~min,
and calculate $\Delta r^2(t_w,\Delta t)$ for each particle
at each time $t_w$.  
The results that follow do not depend on this choice of $\Delta t$.
We then average the mobility over all
particles of a given size with the same number of large or
small neighbors, and plot these results in Fig.~\ref{mob}.
We observe that, on average, having more small neighbors
allows a reference particle to be more mobile.  Conversely,
a local environment rich in large neighbors tends to inhibit
mobility. This seems to hold whether the reference particle is
large or small, as can be seen by comparing the top and bottom
rows of Fig.~\ref{mob}.  Therefore, in this binary sample,
not only are the small particles more mobile on average,
but they also facilitate the mobility of their neighbors
independently of their size.  Of course, this is only true on
average; these results vary quite a lot from particle to particle
\cite{conrad05}.

This is similar to previous observations of binary systems
using rheology \cite{hoffman92,mewis94} and light scattering
\cite{vanmegen01}.  These studies found that binary suspensions
have lower viscosities and faster microscopic motion compared
to monodisperse samples of equal total volume fraction.  This was
attributed to the fact that binary suspensions are capable of
packing to higher volume fractions; thus at a given volume
fraction, there is more free volume for a binary suspension as
compared to a monodisperse suspension.  In particular, a glassy
binary suspension has faster dynamics and can even become
liquid-like with a sufficiently high population of small
particles \cite{vanmegen01}.  For our experiment, we have
purposely chosen the overall volume fraction to ensure that the
sample still has glassy behavior.  Nonetheless, Fig.~\ref{mob} shows
that the
small particles ``lubricate'' the motion of the large ones, as
proposed to interpret the earlier experiments \cite{vanmegen01}.

\begin{figure}[t]
\centering
\includegraphics[width=7.0cm]{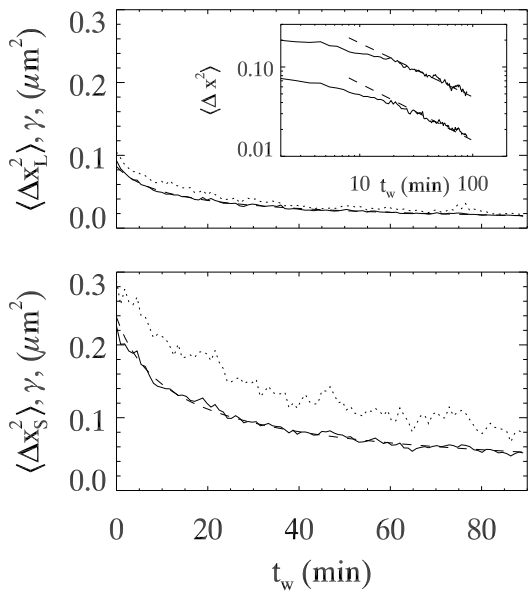}
\caption{$\langle{\Delta x^{2}} \rangle$ (solid) and $\gamma =
  \sqrt{\langle x^{4} \rangle / 3} $ (dotted) as a function of sample
  age for large particles and small particles as
  indicated.  The displacements are defined using $\Delta
  t = 10$ min.  The solid lines are fit to power laws (dashed). 
  Inset:  Same data plotted on a log-log plot; the upper line
  corresponds to the small particles, and the lower line to
  the large particles.
}
\label{x2}
\end{figure}

 
To further investigate the role the small particles have in the aging
of the sample, we examine the distribution of particle
displacements over the same lag time of $10$ minutes.  Figure~\ref{x2}
shows $\langle \Delta x^2(t_w) \rangle$ (solid lines), which is the second
moment of the distribution of displacements, as a function of sample
age, $t_{w}$, for large particles (top) and small particles (bottom).  
Note that here the angle brackets $\langle \cdot \rangle$ indicate an
average over all particles, but not over times $t_w$.  The data
decrease, reflecting the slowing of motion already seen in
Fig.~\ref{msd}.  The small fluctuations in each graph indicate
periods of extra activity (of the sort pictured in
Fig.~\ref{matlab}), and have been seen previously
\cite{CourtlandWeeksJPhysCondMat2003,cipelletti03,cipelletti05,castillo07}.

The non-Gaussianity of the sample can be determined by
comparing these solid curves to the dotted curves, which
represent $\gamma = \sqrt{\langle \Delta x^{4} \rangle / 3}$.
The ratio $\gamma / \langle \Delta x^{2} \rangle = 1$ for a
Gaussian distribution. In our sample, this ratio is $\approx
1.2 $ for the large particles and $\approx 1.6$ for the
small particles.  Both of these ratios are greater than $1$
indicating that the dynamics are anomalous: there is an
excess of large steps when compared to a Gaussian distribution
with the same width. This result is typical in a monodisperse
dense suspension \cite{WeeksWeitzScience2000}. Interestingly,
this ratio is slightly greater for the small particles,
supporting our assertion that the small particles move in
an even more heterogeneous way. This suggests that
the subset of small particles with unusually high mobilities
dominate the dynamics.  Note that we use $\Delta t=10$~min
because the experiment includes time scales $t_w \ll \Delta
t$, $t_w \sim \Delta t$, and $t_w \gg \Delta t$, and the
non-Gaussian behavior of the small particles is present at all of
these time scales.

To characterize the slowing of the dynamics, we fit the large $t_w$
decay of $\langle x^2 \rangle(t_w)$ in Fig.~\ref{x2} to a power
law decay of the form $\langle \Delta x^{2} \rangle \sim t^{-b}$
where $b_{L} =0.64 \pm 0.05$ for the large particles and $b_{S}
= 0.60 \pm 0.05$ for the small particles; these fits are shown
as dashed lines in the figure.  The inset of Fig.~\ref{x2} shows
a log-log plot with the same data and highlights the power-law
decay at large $t_w$.  We note that it is possible and perhaps
likely that this power-law decay is a transient effect that could
disappear at substantially longer $t_w$ \cite{mckenna03}; for our
data, we regard this as simply one way to characterize the behavior.
The values of the exponents $b$ vary from experiment to experiment
and are in the range of 0.60 - 0.80 for this sample.  Within our
uncertainty, the large and small particles decay almost at the
same rates, as suggested by the similar shapes of the MSD curves
(Fig.~\ref{msd}).  However, comparisons between different data
sets always find $b_L > b_S$ by a small amount.  A study of aging
in a monodisperse colloidal sample found a range of power law decay
exponents of $0.05 < b < 0.5$ \cite{CourtlandWeeksJPhysCondMat2003}.
This prior experiment studied colloidal particles with radius $a
\approx 1.18$~$\mu$m which is intermediate to the particle radii
used in our experiment.  However, the power law decay exponent
range found in Ref.~\cite{CourtlandWeeksJPhysCondMat2003} is
less than both exponents found in this study; apparently this
binary suspension ages ``faster''.  Given our observations that
the smaller particles facilitate the motion of their neighbors,
perhaps the presence of these small particles results in this faster
aging as compared with Ref.~\cite{CourtlandWeeksJPhysCondMat2003}.

\section{Conclusion}

We study aging in a binary colloidal glass using confocal
microscopy and distinguish between the small and large
particles within the sample.  These two particle species
slow down with age similarly to each other, but the smaller
particles appear to be more important in the aging process.
These small particles are more mobile, more dynamically
heterogeneous, and facilitate the motion of other small
particles and the motion of the large particles.  Furthermore,
significant motions within the sample (presumably responsible
for the aging) take place in cooperative groups of mobile
particles.  From a practical viewpoint, these results suggest
that the aging of a sample could be influenced by controlling
the size ratio and number ratio of the two species.  Another
likelihood is that a poorly-mixed sample would have spatially
heterogeneous aging, depending on the local composition.

\acknowledgements

We thank E.~Baker, G.~L.~Hunter, and C.~B.~Roth for useful
conversations.  This work was funded by the National Science
Foundation under Grant No.~DMR-0239109.  The work of J.~M.~L.~was
also supported by a SIRE Independent Research Grant from Emory
University.

\end{document}